\documentclass[paper,prd,reprint,preprintnumbers,nofootinbib,notitlepage,showpacs,showkeys]{revtex4-1}
\usepackage{latexsym,graphicx,amssymb,amsmath,mathrsfs} 

\DeclareSymbolFont{bbold}{U}{bbold}{m}{n}
\DeclareSymbolFontAlphabet{\mathbbold}{bbold}

\newcommand{\be}{\begin{equation}}      
\newcommand{\ee}{\end{equation}}      
\newcommand{\bea}{\begin{eqnarray}}      
\newcommand{\eea}{\end{eqnarray}}    
     
\newcommand{\rt}[1]{{}}

\newcommand{\Tr}{\,\textrm{Tr}\,}
\newcommand{\STr}{\,\textrm{STr}\,}
\newcommand{\GeV}{\,\textrm{GeV}\,}
\newcommand{\MeV}{\,\textrm{MeV}\,}
 
\newcommand{\ns}{\,\textrm{ns}\,} 
\newcommand{\s}{\,\textrm{s}\,}

\newcommand{\ife}{\,\textrm{if}\,}
\newcommand{\onePI}{\,\textrm{1PI}\,}
\newcommand{\orr}{\,\textrm{or}\,} 
\newcommand{\els}{\,\textrm{else}\,} 

\makeatletter
\renewcommand\appendix{\par
\setcounter{section}{0}%
\setcounter{subsection}{0}%
\gdef\thesection{\appendixname\space\@Alph\c@section}}

\long\def\unmarkedfootnote#1{{\long\def\@makefntext##1{##1}\footnotetext{#1}}}
\makeatother

\begin{document} 

\title{Thermal properties and evolution of the $U_A(1)$ factor for 2+1 flavors}
\author{G. Fej\H{o}s}
\email{fejos@rcnp.osaka-u.ac.jp}
\author{A. Hosaka}
\email{hosaka@rcnp.osaka-u.ac.jp}
\affiliation{Research Center for Nuclear Physics, Osaka University, Ibaraki, Osaka 567-0047, Japan}

\begin{abstract}
{The thermal evolution of the axial anomaly is investigated in the system of the linear sigma model for $2+1$ flavors. We explore the functional form of the effective potential and the coefficient of the `t Hooft determinant term. It is found that the latter develops a nontrivial structure as a function of the chiral condensate and grows everywhere with respect to the temperature. This shows that mesonic fluctuations strengthen the axial anomaly at finite temperature and it does not vanish at the critical point. The phenomenon has been found to have significance in the thermal properties of the mesonic spectra, especially concerning the $\eta-\eta'$ system.}
\end{abstract}

\maketitle

\section{Introduction}

The fate of the chiral anomaly of quantum chromodynamics (QCD) at finite temperature remains to be understood theoretically. It is well established that at high enough temperature, due to the exponential damping of the instanton density, the anomalous breaking of the $U_A(1)$ subgroup of chiral symmetry has to recover \cite{thooft76,gross81,fukushima13}, but it is virtually unknown what happens around and below the (pseudo)critical temperature ($T_c$). There have been experimental indications that the anomaly might get restored already around the critical temperature of chiral symmetry restoration \cite{vertesi11}, which was also backed by independent theoretical calculations \cite{mitter14,cossu15}, but the issue is not settled as recent lattice approaches (based on the eigenvalue spectrum of the Dirac operator) show a different scenario \cite{dick15,sharma16} and argue that the anomaly is still visible up to $1.5 T_c$.

Earlier, theoretical studies concerning the axial anomaly were performed using effective theories of QCD, in particular the Nambu--Jona-Lasinio model \cite{kunihiro89,fukushima01}, where thermal properties of the quark condensate were included at the mean field level. Effective restoration of the chiral anomaly has also been discussed in nuclear medium in Ref. \cite{nagahiro06}, and is under recent experimental interest at hadron-nuclear facilities \cite{nanova15,LEPS}. In this paper, we focus our attention on the thermal properties of mesons and their effect on the anomaly with mesonic fluctuations taken into account.

Recently, there has been development in the linear sigma model approach, concerning the thermal behavior of the $U_A(1)$ factor. In Ref. \cite{fejos15}, using the so-called chiral invariant expansion technique, it was shown that mesonic fluctuations can have an important contribution in the finite-temperature behavior of the anomaly. They not only produce an anomaly coefficient that is condensate dependent, but depending on region of the parameter space, they can either strengthen or weaken the `t Hooft determinant term, which describes the anomaly in the effective theory approach. Results pointed in a direction where the corresponding determinant coefficient was increasing with the temperature, but no final conclusions could be drawn without proper parametrization of the model, including explicit symmetry-breaking terms representing finite quark masses.

The finite-temperature behavior of the anomaly has consequences on the order of the chiral transition and thus e.g. on the Columbia plot \cite{fukushima11}. In the two-flavor chiral limit (i.e., when quark masses satisfy $m_s \rightarrow \infty$, $m_{u,d} \rightarrow 0$) the presence or absence of the anomaly determines whether an $O(4)$ or $U(2)\times U(2)$ symmetric effective theory describes the phase transition. Traditionally it is argued that the latter produces a first-order transition \cite{pisarski84}, while the former is well known to describe a second-order one. Therefore, the anomaly has an important role in drawing the top-left region of the Columbia plot. We note for completeness that recently there were indications that even in the latter [$U(2)\times U(2)$ model] renormalization group flows can produce an infrared-stable fixed point after all, and thus the transition could be second order anyway \cite{grahl13,pelissetto13,grahl14,nakayama14}, though it might belong to a different universality class than had been thought. This issue still represents an active area of study.

We also mention that there have been finite-temperature investigations in the so-called extended linear sigma model, where scalar mesons are accompanied by vector meson degrees of freedom, the latter being also of importance from a low-energy dynamics point of view. A mean field treatment for $N_f=3$ was presented in Ref. \cite{parganlija13}, and a renormalization group study for $N_f=2$ in Ref. \cite{eser15}.

In this paper we carry out a physical parametrization of the $2+1$-flavor linear sigma model, and calculate numerically the functional form of the effective potential (with the inclusion of explicit breaking terms and the `t Hooft determinant (with a field-dependent coupling). The method we are using is the leading order of the derivative expansion (i.e., local potential approximation) of the functional renormalization group (FRG) approach, combined with the chiral invariant expansion technique \cite{fejos14}, which turned out to reduce numerical costs, allowing the investigatation of a nonperturbative effective potential (i.e., without employing Taylor expansion in terms of couplings) and the corresponding anomaly function. 

The FRG technique has a huge body of literature \cite{wetterich93,berges02,pawlowski07}, and approximate solutions have been very successful in solving low-energy effective theories of QCD in a nonperturbative fashion \cite{skokov10,herbst11,herbst13,herbst14,tripolt14,heller15}. We note, however, that concerning the linear sigma model with $2+1$ flavors and the axial anomaly, there are relatively few works that employ FRG. For earlier studies the reader is referred to Refs. \cite{pawlowski98,jiang12,mitter14,kamikado15,fejos15}. This paper wishes to serve as an improvement and extension along these directions.

The bare coupling of the `t Hooft determinant (being a parameter of the linear sigma model) is in principle determined by QCD dynamics, and thus can be temperature dependent. In this paper, however, we treat this coefficient in a similar fashion as the mass parameter and couplings, and assume that it is given a fixed value and determined by parametrization. We note, however, that, its temperature dependence caused by QCD instantons might be relevant for the fluctuation-corrected anomaly function appearing in the quantum effective action. In other words, the current study investigates how mesonic quantum and thermal fluctuations affect the anomaly, and does not raise questions on the underlying $U_A(1)$ breaking dynamics. Investigations in this direction are beyond the scope of this paper, and remain to be studied in the future.

The paper is organized as follows. In Sec. II we introduce the FRG method, the model, and the corresponding renormalization group flows. Section III is dedicated to explaining the numerics used and some details of the parametrization. The reader can find all numerical results with figures in Sec. IV, while Sec. V is dedicated to conclusions.

\section{Renormalization group flows}

A powerful way to incorporate quantum and thermal fluctuations in quantum field theories is the application of the FRG method. It generalizes the concept of the conventional Wilsonian RG in the sense that instead of deriving flow equations for individual coupling constants, one has an exact evolution equation for the effective action itself. Through a momentum-dependent mass term -- which is defined with the help of an IR regulator function $R_k$ where $k$ plays the role of the flow parameter -- one suppresses modes with momenta $q \lesssim k$, and gradually integrates them out by taking the $k\rightarrow 0$ limit. This idea leads to a one-parameter family of quantum effective actions ($\Gamma_k$) obeying the so-called Wetterich equation \cite{wetterich93}:
\bea
\label{Eq:Wet}
\partial_k \Gamma_k = \frac12 \STr \left[ (\Gamma_k^{(2)}+R_k)^{-1} \partial_k R_k\right],
\eea
where $\Gamma_k^{(2)}$ is the second functional derivative of $\Gamma_k$ and the STr operation has to be taken in both the functional and matrix sense. If $k$ is large (practically equal to a UV cutoff $\Lambda$), no fluctuations are included in $\Gamma_k$ and it takes the form of the classical action $S$. On the other hand, at $k=0$ the IR regulator vanishes and one obtains the usual one-particle-irreducible (1PI) effective action, $\Gamma_{k=0}=\Gamma_{\onePI}$. In practice, one considers a theory at the highest scale $\Lambda$ and sets up the RG flow initial condition as the classical expression for the action, and integrates it down to $k=0$. It has to be noted that the evolution equation (\ref{Eq:Wet}) is an exact relation and has to be approximated for practical purposes.

The effective model of the strong interaction we employ here is the three-flavor linear sigma model. In accordance with Lorentz symmetry and renormalizability, we assume that at the UV scale the action is of the following form:
\bea
\label{Eq:S}
S&=&\int_x \Big(\Tr(\partial_\mu M^\dagger \partial^\mu M) - \mu^2 \Tr (M^\dagger M) \nonumber\\
&-& \frac{g_1}{9} \left[\Tr(M^\dagger M)\right]^2 - \frac{g_2}{3} \Tr (M^\dagger M M^\dagger M)\nonumber\\
&-&\Tr\big(H(M^\dagger+M)\big)-a (\det M^\dagger + \det M)\Big),
\eea
where $M=T^b(\sigma^b+i \pi^b)$ is a $3\times 3$ matrix [element of the $U(3)$ Lie algebra generated by $T^b$, $\Tr (T^bT^c)=\delta^{bc}/2$], playing the role of the order parameter of chiral symmetry breaking. Its fluctuations correspond to scalar and pseudoscalar mesons $a_0, \kappa, f_0, \sigma$ and $\pi, K, \eta, \eta'$, respectively. In (\ref{Eq:S}), terms in the first two lines are invariant under $M \rightarrow LMR^\dagger$ chiral symmetry (where $L$ and $R$ are unitary matrices with parameters $\theta^b_{L/R}$), the fourth term breaks it explicitly, while the last one corresponds to the $U_A(1)$ anomaly (i.e., `t Hooft determinant). Note that the chiral transformations can also be expressed in terms of vector- and axialvector transformations, with the transformation parameters $\theta^b_{V/A}=(\theta^b_L\pm \theta^b_R)/2$. Since we are describing spontaneous breaking of chiral symmetry, the mass parameter is negative, $\mu^2<0$, and the stability conditions imply $g_1+g_2>0$, $g_2>0$ \cite{fejos13}. We choose the explicit breaking matrix $H$ to be $H=h_0T^0+h_8 T^8$ (no isospin breaking), which leads to
\bea
\Tr\big(H(M^\dagger+M)\big)=h_0s^0+h_8s^8.
\eea
As announced already, $\Gamma_k$ has to be approximated to solve Eq. (\ref{Eq:Wet}). We choose to employ the derivative expansion at leading order (this is also called the local potential approximation),
\begin{subequations}
\label{Eq:Ansatz}
\bea
\label{Eq:Ansatzgamma}
\Gamma_k&=& \int_x \Big(\Tr(\partial_\mu M^\dagger \partial^\mu M)-V_k(M)\Big),\\
\label{Eq:AnsatzV}
V_k&=&U_k(I_1)+C_k(I_1)I_2+(h_0s^0+h_8s^8) \nonumber\\
&&+A_k(I_1) I_{\det}, 
\eea
\end{subequations}
where we have neglected the wave-function renormalization and introduced the following chiral invariants:
\bea
I_1&=&\Tr (M^\dagger M), \nonumber\\
I_2&=&\Tr (M^\dagger M - \Tr(M^\dagger M)/3)^2,
\eea
and the `t Hooft determinant
\bea
I_{\det} &=& \det M^\dagger + \det M.
\eea
The latter only breaks the $U_A(1)$ subgroup, and note that in the ansatz (\ref{Eq:Ansatzgamma})--(\ref{Eq:AnsatzV}) we have left out higher order $U_A(1)$ breaking Lorentz scalar operators, such as $(\det M^\dagger + \det M)^i$, $i\geq 2$, and $(\det M^\dagger - \det M)^{2j}$, $j\geq 1$, even though they are in principle generated by Eq. (\ref{Eq:Wet}). Matching Eqs. (\ref{Eq:Ansatzgamma})--(\ref{Eq:AnsatzV}) with Eq. (\ref{Eq:S}) shows 
\bea
U_\Lambda=\mu^2I_1 + \frac{g_1+g_2}{9} I_1^2, \quad C_\Lambda=\frac{g_2}{3}, \quad A_\Lambda=a.
\eea
In obtaining the specific form of Eq. (\ref{Eq:AnsatzV}), we have also made use of the chiral invariant expansion technique which exploits the fact that -- besides the explicit symmetry breaking terms -- the effective action has to be invariant under chiral symmetry. The reader is referred for details to Ref. \cite{fejos14}. Note that the explicit breaking of chiral symmetry via $\Tr\left(H(M^\dagger+M)\right)$ does not change with respect to the RG flow. This is due to the fact that on the right-hand side of Eq. (\ref{Eq:Wet}) the $\Gamma_k$ function enters only via its second functional derivative and thus terms linear in the fields do contribute to the flow equation. As a consequence, there is no way to generate any kind of explicit symmetry breaking terms on the right-hand side of Eq. (\ref{Eq:Wet}) at a given scale $k$, and therefore they are uniquely determined by the UV action through parametrization.

If one plugs Eq. (\ref{Eq:Ansatzgamma}) into Eq. (\ref{Eq:Wet}), using Litim's optimal regulator \cite{litim01}, $R_k(q_0,{\bf q})=(k^2-{\bf q}^2)\Theta(k^2-{\bf q}^2)$, and evaluating the loop integral at finite temperature one gets the following evolution equation for the $k$-dependent effective potential:
\bea
\label{Eq:effpotflow}
\partial_k V_k = \frac{k^4T}{6\pi^2}\sum_{j=-\infty}^\infty\sum_{\alpha=s,\pi}\sum_{i=0}^8 \frac{1}{\omega_j^2+k^2+m_{\alpha,i}^2(k)},
\\ \nonumber
\eea
where $\omega_j=2\pi j T$ are bosonic Matsubara frequencies, and $m_{s/\pi,i}^2(k)$ denote the eigenvalues of the scalar and pseudoscalar mass matrices (defined as $\partial^2 V_k/\partial s_i\partial s_j$ and $\partial^2 V_k/\partial \pi_i\partial \pi_j$), respectively, evaluated at scale $k$. Note that the applied regulator is three dimensional in the sense that it has no cutoff in the timelike direction ($q_0$), leading the Matsubara sums to run all over the possible values of $\omega_j$.

The way to translate Eq. (\ref{Eq:effpotflow}) into the flows of $U_k$, $C_k$, and $A_k$ is the following. First we set the anomaly to zero and consider a background field that consists of two independent condensates. Using the notation $\bar{s}_i=v_i$ for constant mean fields, we write $\bar{M}=v_0T^0+v_8T^8$. In this two-component background, the invariants take the following form:
\vspace{1cm}
\bea
I_1&=&\frac{v_0^2+v_8^2}{2}, \quad I_2=\frac{v_8^2}{24}(v_8-2\sqrt{2}v_0)^2, \nonumber\\
I_{\det}&=&\frac{1}{3\sqrt6}(v_0^3-\frac32v_0v_8^2-\frac{1}{\sqrt2}v_8^3), \\ \nonumber
\eea
and their field derivatives that are necessary for calculating the masses can be found in the Appendix. After calculating the right-hand side of Eq. (\ref{Eq:effpotflow}) in this background field, it is easy to combine terms into the invariant tensors above (expanding the expression around $v_8 \approx 0$ helps) and thus first one identifies the flow of $U_k(I_1)$, and then the coefficient of $I_2$, which is nothing but the flow of $C_k(I_1)$. As a second step, one includes the anomaly and Taylor expand the right-hand side of Eq. (\ref{Eq:effpotflow}) around the zero anomaly configuration up to next-to-leading order. This leads to the appearance of the `t Hooft determinant $I_{\det}$ and its coefficient provides the flow of $A_k(I_1)$. The obtained equations are \cite{fejos15}
\begin{widetext}
\begin{subequations}
\bea
\label{Eq:flow_Uk}
\partial_kU_k(I_1)&=&\frac{k^4T}{6\pi^2}\sum_{j=-\infty}^\infty\Bigg[\frac{9}{\omega_j^2+E_\pi^2}+\frac{8}{\omega_j^2+E_{a_0}^2}+\frac{1}{\omega_j^2+E_\sigma^2}\Bigg], \\
\label{Eq:flow_Ck}
\partial_k C_k(I_1)&=&\frac{k^4T}{6\pi^2}\sum_{j=-\infty}^\infty\Bigg[\frac{4(3C_k+2I_1C_k')^2/3}{(\omega_j^2+E_{a_0}^2)^2(\omega_j^2+E_\sigma^2)}
+\frac{128C_k^5I_1^3/3}{(\omega_j^2+E_\pi^2)^3(\omega_j^2+E_{a_0}^2)^3}+\frac{24C_k\left(C_k-I_1C_k'\right)}{(\omega_j^2+E_{a_0}^2)^3}\nonumber\\
&+&\frac{4\left(3C_kC_k'I_1+4I_1^2C_k'+C_k(3C_k-2C_k''I_1^2)\right)/3}{(\omega_j^2+E_{a_0}^2)(\omega_j^2+E_\sigma^2)^2}
+\frac{64C_k^3I_1^2(C_k-I_1C_k')/3}{(\omega_j^2+E_\pi^2)^2(\omega_j^2+E_{a_0}^2)^3}-\frac{48C_k^2I_1^2C_k'}{(\omega_j^2+E_\pi^2)(\omega_j^2+E_{a_0}^2)^3} \nonumber\\
&+&\frac{6C_k-17I_1C_k'}{(\omega_j^2+E_{a_0}^2)^2}\frac{1}{I_1}-\frac{6C_k+9I_1C_k'+2I_1^2C_k''}{(\omega_j^2+E_\sigma^2)^2}\frac{1}{I_1}
+\frac{4C_k(6C_k+9I_1C_k'+2I_1^2C_k'')/3}{(\omega_j^2+E_{a_0}^2)(\omega_j^2+E_\sigma^2)^2}\Bigg], \\
\label{Eq:flow_Ak}
\partial_k A_k(I_1)&=&\frac{k^4T}{6\pi^2}\sum_{j=-\infty}^\infty\Bigg[-\frac{9A_k'}{(\omega_j^2+E_\pi^2)^2}-\frac{9A_k}{I_1(\omega_j^2+E_\pi^2)^2}
-\frac{8A_k'}{(\omega_j^2+E_{a_0}^2)^2}+\frac{12A_k}{I_1(\omega_j^2+E_{a_0}^2)^2} \nonumber\\
&-&\frac{3A_k}{(\omega_j^2+E_\sigma^2)^2I_1}+\frac{7A_k'}{(\omega_j^2+E_\sigma^2)^2}+\frac{2I_1A_k''}{(\omega_j^2+E_\sigma^2)^2}\Bigg]. \\ \nonumber
\eea
\end{subequations}
\end{widetext}
Here we have introduced the following shorthand notations: $E_\pi^2=k^2+U_k'(I_1)$, $E_{a_0}^2=k^2+U_k'(I_1)+\frac43 I_1 C_k(I_1)$,
$E_{\sigma}^2=k^2+U_k'(I_1)+2I_1U_k''(I_1)$, which are energies of the $\pi$, $a_0$ and $\sigma$ mesons, respectively, in a background satisfying $I_1\neq 0$, $I_2=0$ (e.g. $v_0\neq 0$, $v_8=0$). We also note that all Matsubara sums can be performed analytically, thus the integro-differential nature of the Wetterich equation reduces to coupled (functional) differential equations of the $U_k$, $C_k$, $A_k$ coefficient functions.

\section{Numerics and parametrization}

The coupled equations (\ref{Eq:flow_Uk}), (\ref{Eq:flow_Ck}), and (\ref{Eq:flow_Ak}) are solved by the grid method. Calculations are carried out in GeV units, which is also the UV cutoff $\Lambda$, from which the equation system is integrated. We set up one-dimensional grids in the interval $[0$:$2]$ with a step size of $10^{-2}$ (i.e., $10$ MeV) and solve the field equations at each point. Field derivatives are calculated with the seven-point formula, except close to the boundaries, where the five- and three-point formulas are used. Using suitable initial values for $\mu^2$, $g_1$, $g_2$, $a$, $h_0$, $h_8$ (note again that the latter two do not change with $k$) we integrate the functions from $k=\Lambda$ down to $k=0$, using the Runge-Kutta algorithm.

\begin{table}[t]
\centering
\vspace{0.2cm}
  \begin{tabular}{ c | c }
    particle & mass [MeV] \\ \hline
    $\pi$ & $139.57$ \\ \hline
    K & $493.68 \pm 0.02$ \\ \hline
    $\eta$ & $547.86 \pm 0.02$ \\ \hline
    $\eta'$ & $957.78 \pm 0.06$ \\ \hline
    $a_0$ & $980 \pm 20$ \\ \hline
    $\kappa$ & $682 \pm 29$ \\ \hline
    $f_0$ & $990 \pm 20$ \\ \hline
    $\sigma$ & $400$-$550$ \\ 
  \end{tabular}
  \caption{Zero-temperature mass spectrum based on the Particle Data Group \cite{pdg}. $\pi$, $K$, $\eta$, and $\eta'$ are used for parametrization for the nonzero anomaly case. For zero anomaly, the $\eta-\eta'$ system is replaced by $\sigma$.}
\end{table}
We note that the anomaly coefficient $a$ defined in Eq. (\ref{Eq:S}) has to be negative for physical parametrizations. This stems from the fact that in the case when no explicit symmetry breaking terms are present, it is easy to show that the pseudoscalar spectrum takes the following form at the minimum point of the effective potential:
\bea
m_{\pi,0}^2&=&\frac{3a(3a-\sqrt{-24\mu^2(g_1+g_2)+9a^2})}{4(g_1+g_2)}, \nonumber\\
m^2_{\pi,i}&=&0 \quad (i=1,...8),
\eea
where $m_{\pi,0}^2$ corresponds to the $\eta'$ meson and it is positive if and only if $a<0$. This is expected to remain true when the explicit breaking terms are introduced, and also after fluctuations are included. It will be confirmed by numerics in the next section.

First the values of $h_0$ and $h_8$ are fixed. We use the partially conserved axial-vector current (PCAC) relations, i.e., \cite{peskin}
\bea
\label{Eq:pcac}
\partial_\mu J^{\mu 5}_b = m_{\pi,b}^2 f_b \pi_b \qquad (b=0,...8),
\eea
where $J_b^{\mu 5}$ is the flavor nonet axial-vector current, and $f_b$ denotes corresponding decay constants. In the case when explicit symmetry breaking terms are present, the divergence of the current is explicitly given by
\bea
\label{Eq:axdiv}
\partial_\mu J^{\mu 5}_b = -\frac{\partial}{\partial \theta_A^b}\Tr \Big(H(M+M^\dagger)\Big).
\eea
Calculating the right-hand side of Eq. (\ref{Eq:axdiv}) and then combining Eq. (\ref{Eq:pcac}) with Eq. (\ref{Eq:axdiv}), we get the following, separately for $b=1,2,3$ and $b=4,5,6,7$:
\begin{subequations}
\bea
m_\pi^2 f_\pi &=& \sqrt{\frac{2}{3}}h_0+\frac{1}{\sqrt3}h_8, \\
m_K^2 f_K &=&\sqrt{\frac{2}{3}}h_0-\frac12\frac{1}{\sqrt3}h_8,
\eea
\end{subequations}
where now $m_\pi$ and $m_K$ denote the physical pion and kaon masses, respectively, and $f_\pi$, $f_K$ are decay constants, for which we employ $f_\pi \approx 93$ MeV and $f_K \approx 113$ MeV. This determines the explicit breaking parameters:
\bea
h_0 = (286 \MeV)^3, \quad h_8 = -(311 \MeV)^3.
\eea
For the remaining four undetermined parameters (i.e., $\mu^2, g_1, g_2, a$) we use the $\pi, K, \eta, \eta'$ mass values listed in Table I. Solving Eqs. (\ref{Eq:flow_Uk}), (\ref{Eq:flow_Ck}) and (\ref{Eq:flow_Ak}) in a way that reproduces these masses at zero temperature selects an appropriate choice of parameters at the UV scale. Note that if we set $a=0$ in advance, obviously the physical value of the $\eta'$ mass cannot be maintained, and also $\eta$ gets degenerated with the pions. In this case, besides the pion and kaon masses, we used the $\sigma$ mass for parametrization. After all parameters are fixed, one is free to solve the equations at any given temperature, at which the ground state and the mass spectrum will be predictions. A list of the obtained values can be found in Table II.

\begin{table}[t]
\centering
\vspace{0.2cm}
  \begin{tabular}{ c | c | c }
    parameter & anomaly & no anomaly \\ \hline
    $\mu^2$ & $-0.95 \GeV^2$ & $-0.95 \GeV^2$ \\ \hline
    $g_1$ & 80 & 90 \\ \hline
    $g_2$ & 140 & 200 \\ \hline
    $a$ & $-3.0 \GeV$ & $-$ \\ \hline  
  \end{tabular}
  \caption{Set of parameters at the UV scale used for finite-temperature calculations.}
\end{table}

For the scalar sector, it is convenient to work in the ideal mixing basis [s: strange; ns: nonstrange, see Eqs. (\ref{Eq:sns}) and (\ref{Eq:snsh})] rather than the flavor one. They are related by the transformation
\bea
\label{Eq:sns}
\begin{pmatrix}
\sigma_{\ns} \\ \sigma_{\s}
\end{pmatrix}
= \frac{1}{\sqrt3} \begin{pmatrix}
\sqrt2 & 1 \\
1 & -\sqrt2
\end{pmatrix}
\begin{pmatrix}
\sigma_0 \\ \sigma_8
\end{pmatrix},
\eea
and similarly for the explicit symmetry breaking parameters:
\bea
\label{Eq:snsh}
\begin{pmatrix}
h_{\ns} \\ h_{\s}
\end{pmatrix}
= \frac{1}{\sqrt3} \begin{pmatrix}
\sqrt2 & 1 \\
1 & -\sqrt2
\end{pmatrix}
\begin{pmatrix}
h_0 \\ h_8
\end{pmatrix}.
\eea
This leads to
\bea
h_{\ns} = (121 \MeV)^3 , \quad h_{\s} = (336 \MeV)^3,
\eea
in agreement with Ref. \cite{mitter14}. Once the functions $U_{k\rightarrow 0}(I_1)$, $C_{k\rightarrow 0}(I_1)$,  and  $A_{k\rightarrow 0}(I_1)$ are determined, the full effective potential  $V_{k\rightarrow 0}(I_1)$ is also in our possession. Its minimum determines the ground state of the system, which is obtained by simply taking the field derivatives (numerically) with respect to $v_0$ and $v_8$ and finding their zeros; see also the Appendix.

\begin{figure}
\includegraphics[bb = 50 80 520 810,scale=0.35,angle=-90]{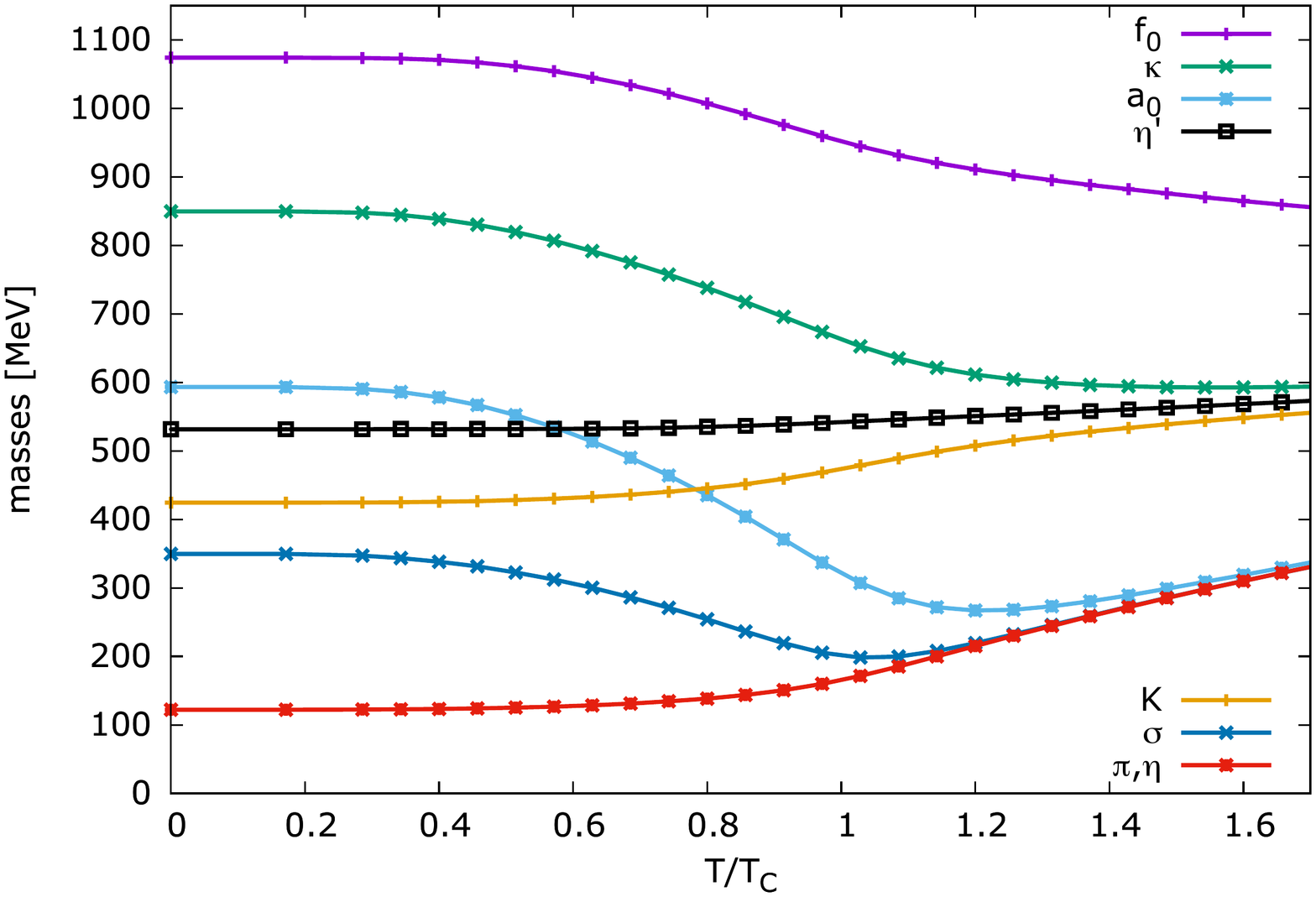}
\caption{Mass spectrum for the zero anomaly case. Parametrization was carried out by fixing the $\pi$, $K$, and $\sigma$ masses. Pions are degenerated with the $\eta$.}
\end{figure}

\setcounter{figure}{2}
\begin{figure}
\includegraphics[bb = 50 80 520 810,scale=0.35,angle=-90]{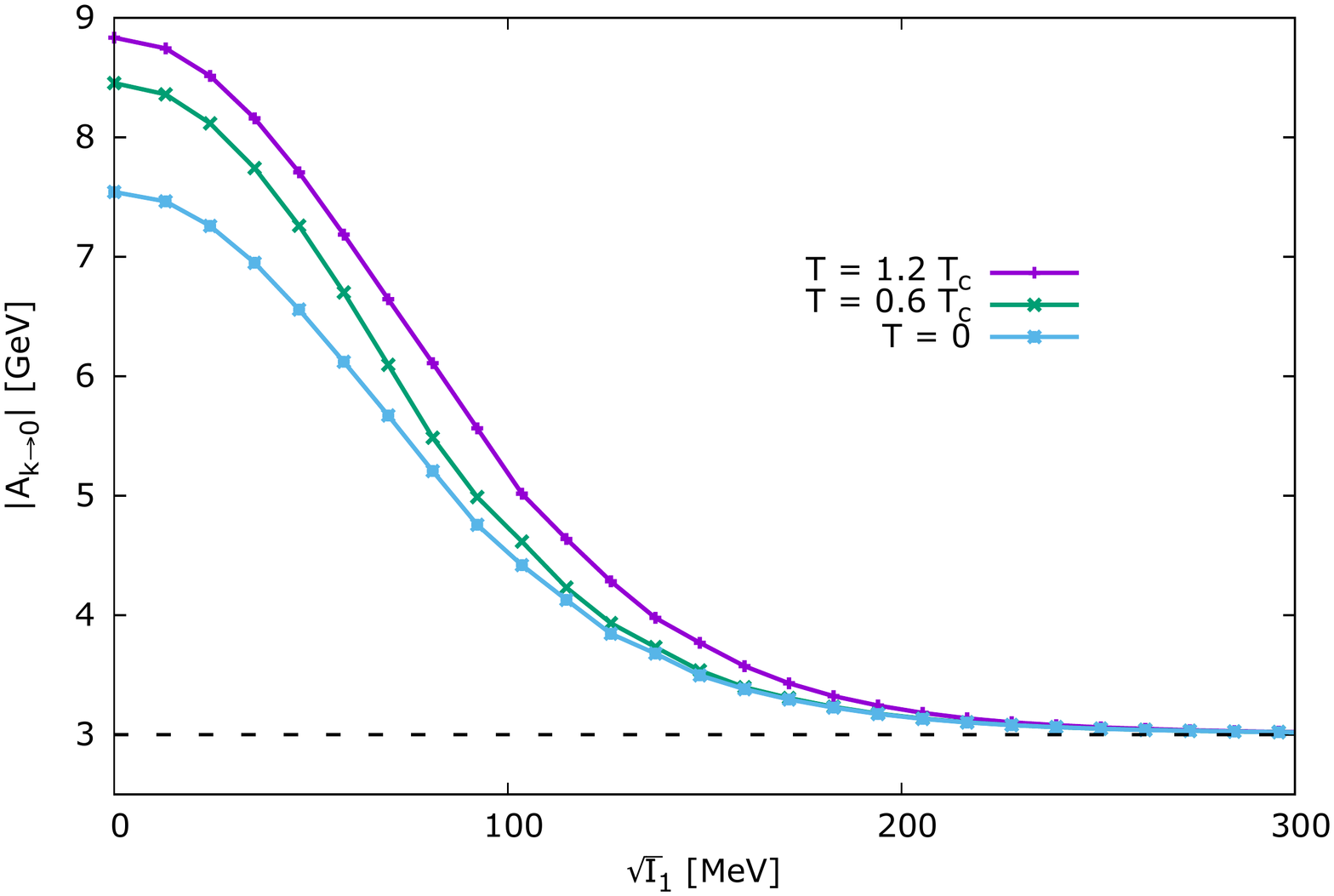}
\caption{Absolute value of the anomaly function $A_{k\rightarrow 0}$ as a function of $\sqrt{I_1}$ [note that $I_1 \equiv (v_0^2+v_8^2)/2$]. The function increases everywhere with respect to the temperature, and deviates significantly from its value in the UV (dashed line).}
\end{figure}

\setcounter{figure}{1}
\begin{figure}
\includegraphics[bb = 50 80 520 810,scale=0.35,angle=-90]{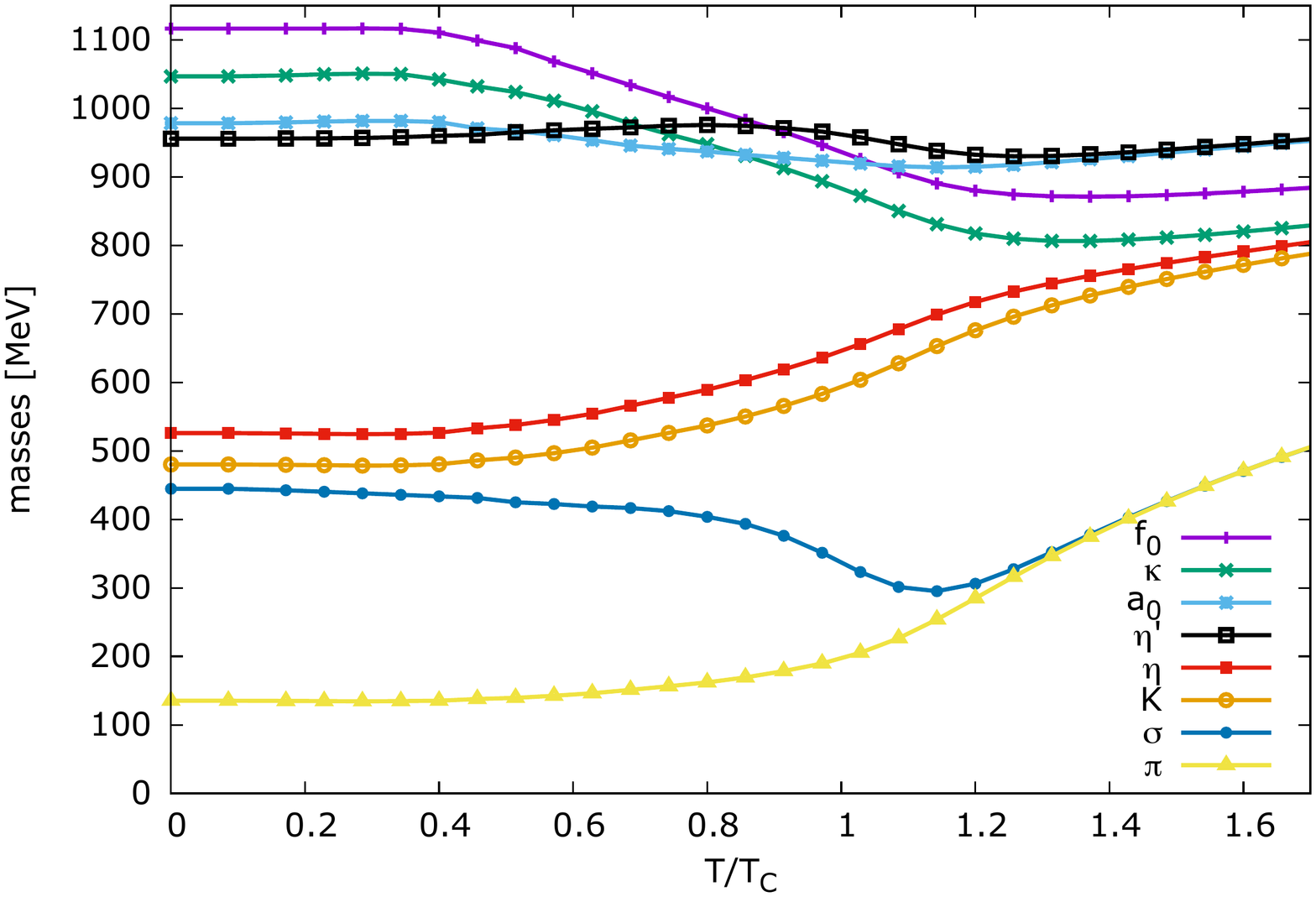}
\caption{Mass spectrum including the $U_A(1)$ anomaly. Parametrization was carried out by fixing the $\pi$, $K$, $\eta$, and $\eta'$ masses.}
\end{figure}

\setcounter{figure}{3}
\begin{figure}
\includegraphics[bb = 50 80 520 810,scale=0.35,angle=-90]{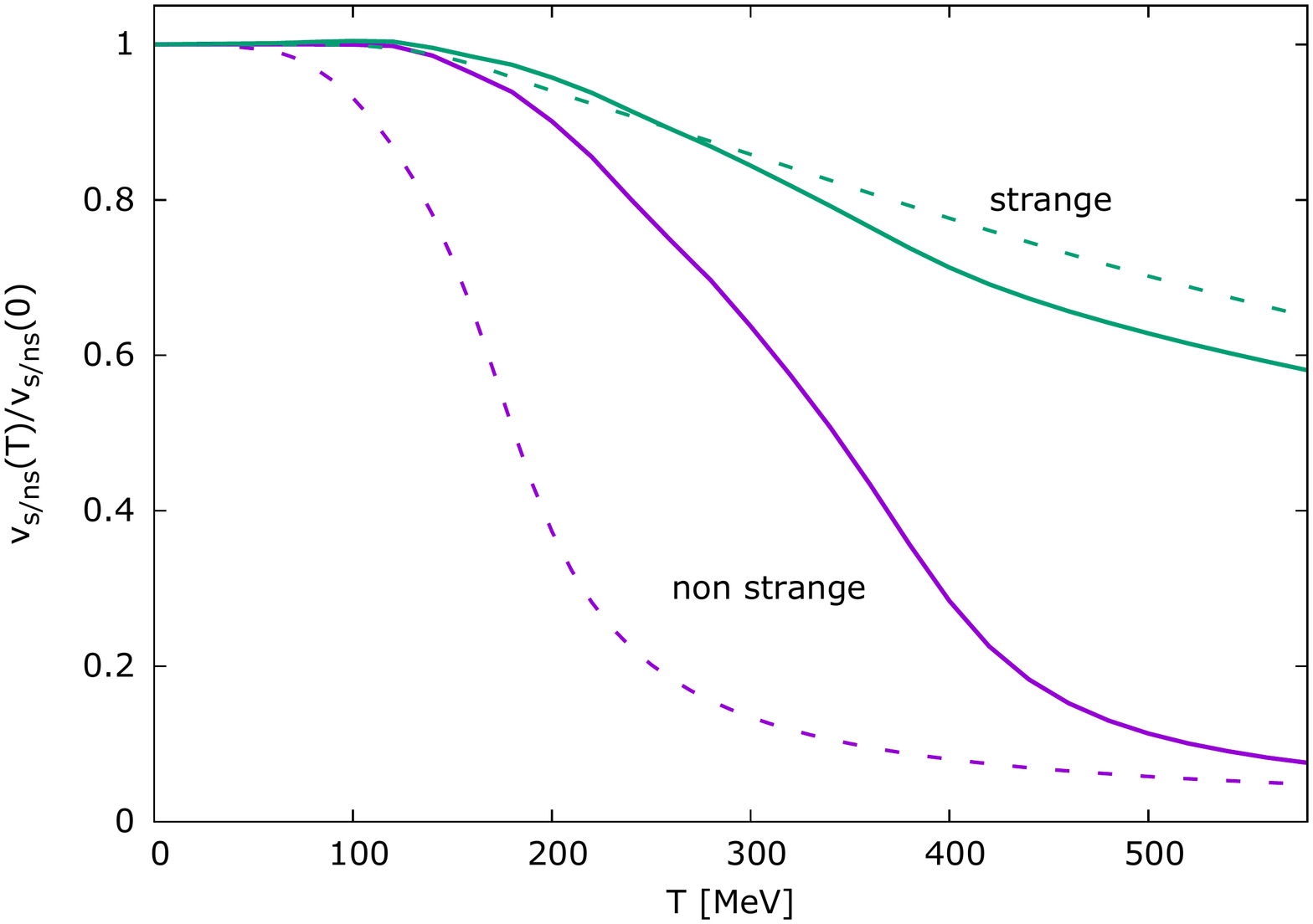}
\caption{Phase transition: Thermal evolution of the strange and nonstrange condensates. The dashed (solid) lines correspond to the zero (finite) anomaly case.}
\end{figure}

\section{Results}

In this section we discuss the results that can be extracted from solving the coupled equation system for $U_k$, $C_k$, and $A_k$. First of all, Fig. 1 and Fig. 2 show the mass spectrum as a function of the temperature (in units of the corresponding critical temperature) for the zero and nonzero anomaly case, respectively. At $T = 0$ and without anomaly (Fig. 1), even if one tries to fix the parameters using the masses of the lightest mesons ($\pi$, $K$, and $\sigma$) as inputs, one cannot reproduce the physical spectrum very accurately. In particular, pions and the $\eta$ meson are degenerate due to the ideal mixing associated with the $SU(3)$ flavor breaking, $m_u = m_d < m_s$. Turning on the anomaly (Fig. 2), however, the spectrum improves significantly, and due to large $U_A(1)$ breaking the physical $\eta$ and $\eta'$ get closer to the $\pi_0\equiv \eta_0$ and $\pi_8\equiv \eta_8$ pseudoscalars, respectively. Generally, the model achieves fair agreement with experimental data within an uncertainty of the order of $10\%$, except for the $\kappa$ meson. Our prediction for its mass lies slightly above $1 \GeV$, which shows that a better identification would probably be the $K_0^*(1430)$ meson. This points in the direction that the conventional identification of light scalar mesons with excitations in the linear sigma model might not be accurate, and some of these particles are predominantly four-quark states \cite{parganlija13}. Nevertheless, e.g. the least known $\sigma$ mass is found to be around $450 \MeV$, consistent with the Particle Data Group \cite{pdg}. We also note that the patterns of the spectra at high temperature corresponds to symmetry breakings where the nonstrange condensate has already evaporated, but the strange one still carries significant contribution.

\begin{figure}
\includegraphics[bb = 50 80 520 810,scale=0.35,angle=-90]{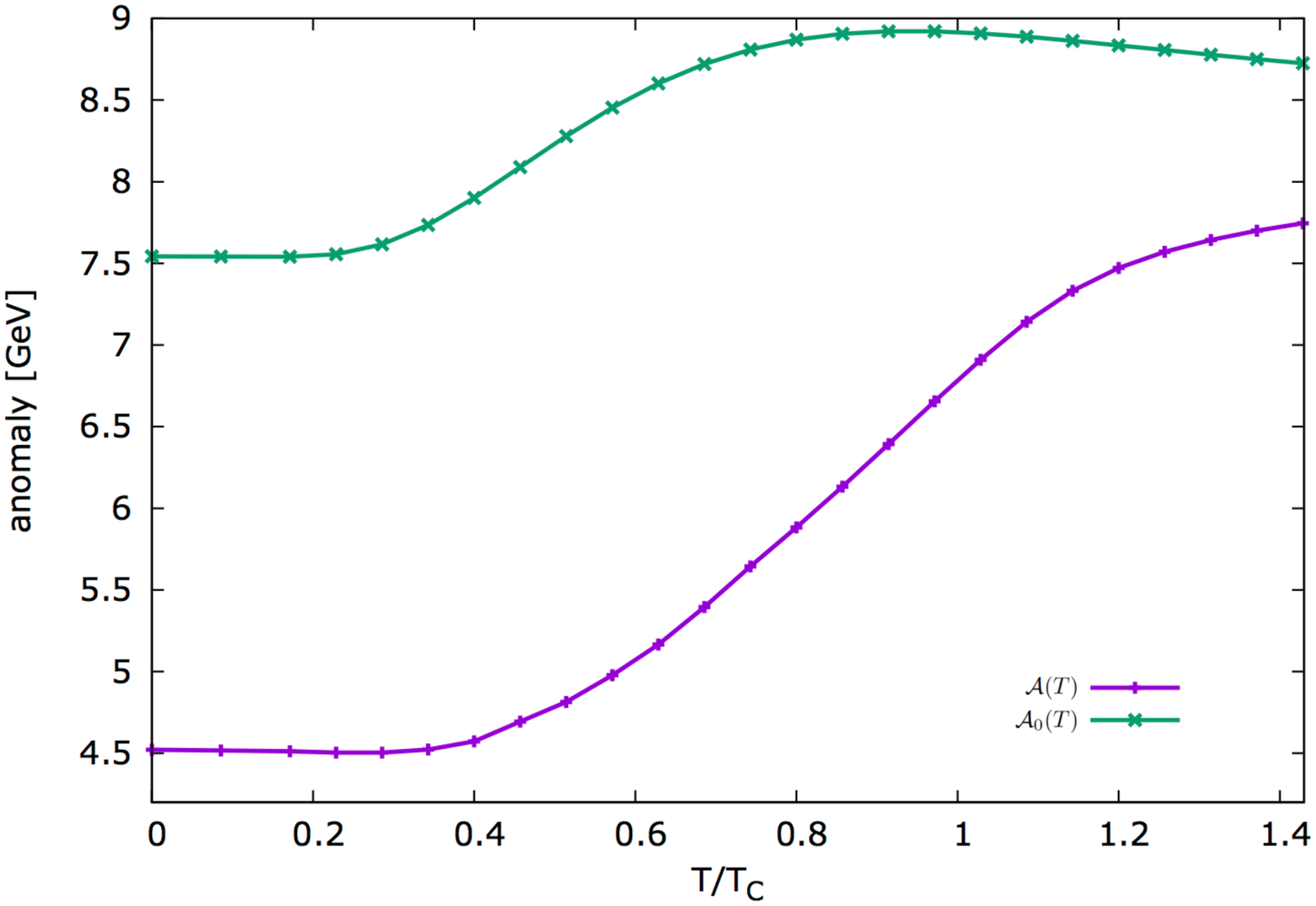}
\caption{Absolute value of the anomaly at zero condensate and at the real ground state as a function of the temperature.}
\end{figure}

Turning to finite temperature, an important observation is that (unlike predictions of earlier works) we observe no drop in the $\eta'$ mass around $T_c$. This points in the direction that the $U_A(1)$ factor might not even be partially restored toward the critical temperature. One still observes a reduction in the $\eta$ and $\eta'$ mass difference, and thus the safest way to draw conclusions on the issue is to explicitly check the temperature dependence of the fluctuation-corrected anomaly function. Figure 3 shows the anomaly function $|A_{k\rightarrow 0}(I_1)|$ as the temperature increases. As suggested by the mass spectrum, the anomaly strengthens everywhere with the temperature, and furthermore it produces a nontrivial structure as a function of the condensates. This clearly reveals the fate of the $U_A(1)$ factor, and predicts that mesonic quantum and thermal fluctuations are indeed of importance, concerning on the one hand the condensate dependence of the `t Hooft coupling, and on the other hand its thermal evolution.

\begin{figure}
\includegraphics[bb = 50 80 520 810,scale=0.35,angle=-90]{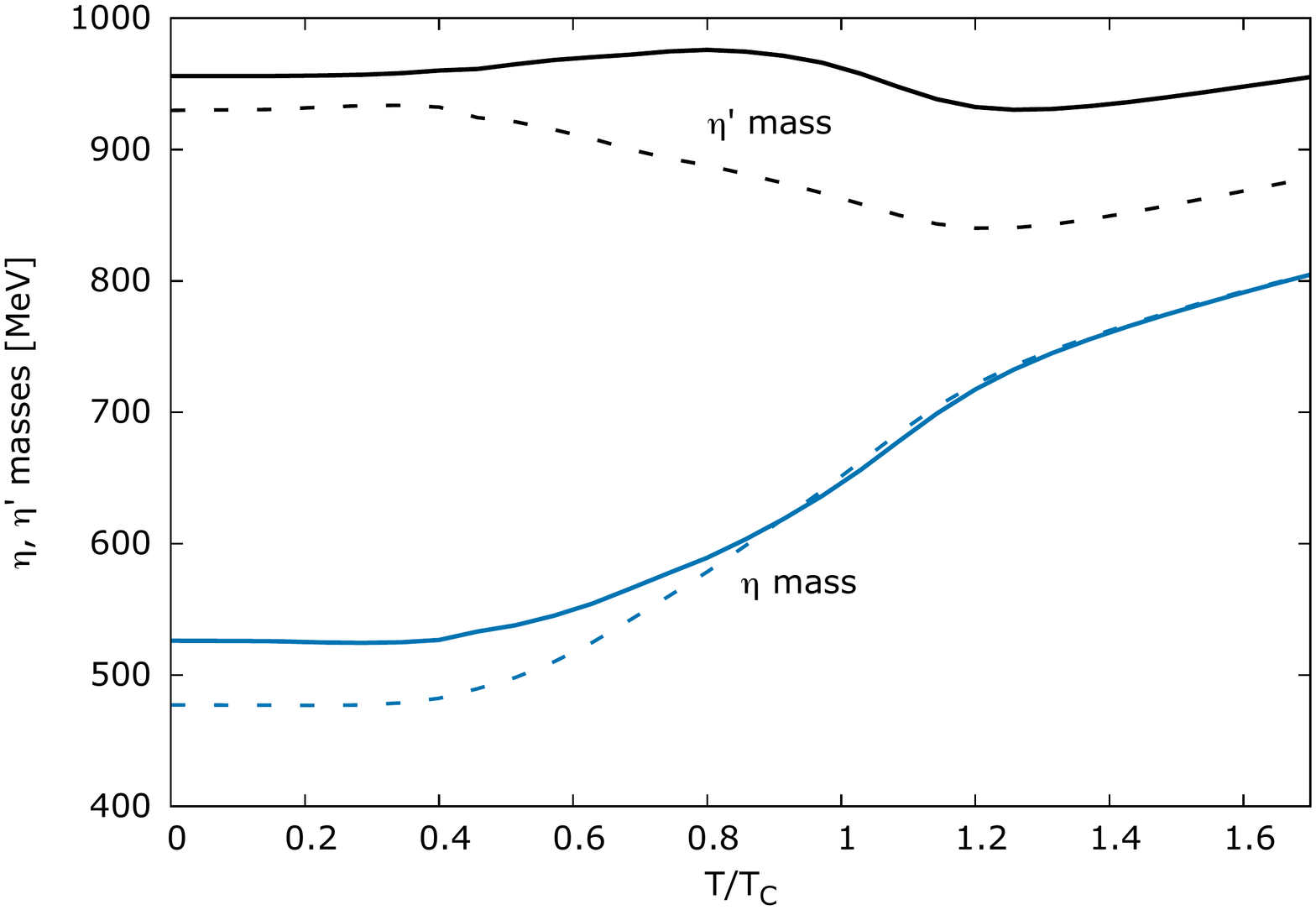}
\caption{$\eta$ and $\eta'$ masses as a function of the temperature. Solid (dashed) lines correspond to the temperature- and field-dependent (independent) anomaly.}
\end{figure}

The phase transition itself can be seen in Fig. 4. We plot both the strange and nonstrange condensates normalized to their $T=0$ value. The pseudocritical temperature of the transition is defined as the inflection point of the non-strange condensate. If the anomaly is turned off, we get $T_c \approx 175$ MeV, which is in decent agreement with lattice data; however, when the anomaly is turned on, it goes up to $T_c \approx 350$ MeV. This clearly signals that the current setup is not capable of being extended toward the critical temperature. It might be related to the fact that even though the transition temperature should be independent of the number of colors, in the linear sigma model one has $T_c \propto \sqrt{N_c}$ \cite{heinz12}, which is not cured by the FRG method. Another possibility is that the high value for $T_c$ is caused by the absence of quark degrees of freedom and/or gauge effects, and by introducing the former through Dirac fermions and the latter via the Polyakov loop, $T_c$ is pursued to a lower value. In spite of all these, we see no obstacle in drawing qualitative conclusions on mesonic fluctuation effects on the $U_A(1)$ anomaly, but it is, however, important to note that there are indications that mesonic fluctuations can be considerably smaller than quark and gauge contributions \cite{herbst14}.

It is also worth noting that the actual strength of the anomaly in the ground state of the system has both explicit and implicit temperature dependence. The explicit dependence comes from the change of the shape of the $A_{k\rightarrow 0}(I_1)$ function, and the implicit one arises from the fact that as the temperature rises, the $I_1$ value corresponding to the ground state changes. One may define two anomaly strengths as
\bea
{\cal A}(T)&=&A_{k\rightarrow 0}(I_1(T);T), \nonumber\\
{\cal A}_0(T)&=&A_{k\rightarrow 0}(0;T),
\eea
which are shown in Fig. 5. With this one can keep track of how explicit and full temperature dependences contribute to the $U_A(1)$ factor.

Finally, let us come back to the $\eta-\eta'$ system. We noted already that we observe no drop in the $\eta'$ mass around the critical temperature. In Fig. 6 it is shown that this is indeed due to the field and temperature dependence of the anomaly function. We plot both the $\eta$ and $\eta'$ masses with the inclusion of the temperature-dependent $A_{k\rightarrow 0}(I_1)$ function, together with a hybrid method, where the anomaly is set to a temperature- and field-independent constant arising from parametrization (i.e., the value of the anomaly coefficient in the ground state at $T=0$). One observes that without thermal effects of the anomaly, we indeed get a drop in the the $\eta'$ mass, as predicted by other methods \cite{mitter14}. We argue, however, that this drop is smoothened by mesonic fluctuations, and we also note that no such effect is observed in the $\eta$ mass. Its thermal behavior is not sensitive to the anomaly treatment, at least qualitatively. One might be interested in why the $T=0$ masses differ in the two cases, and the reason is that the field derivatives of the anomaly also carry non-negligible contributions, which are of course not present in the constant anomaly (hybrid) scenario.

\section{Conclusions}

In this paper we have investigated the quantum and thermal behavior of mesonic fluctuations in the $2+1$-flavor linear sigma model, including their effect on the chiral anomaly. A physical parametrization of the leading order of the derivative expansion in the functional renormalization group formalism has been performed in order to obtain the effective potential of the model leading to the vacuum structure, particle masses, anomaly function and their thermal properties.

We have found that the coefficient of the `t Hooft determinant becomes field dependent, and increases with the temperature, or in other words, mesonic fluctuations strengthen the anomaly toward $T_c$. It also turned out that these effects lead to the smoothening of the temperature dependence of the $\eta'$ mass, and no drop can be observed anymore around the critical temperature. These are in agreement with recent lattice simulations \cite{dick15,sharma16}, which argue that the anomaly does not restore until around $1.5 T_c$. It remains an important question whether the temperature dependence of the `t Hooft parameter that arises from instanton effects can compete with mesonic fluctuations.

If the anomaly is turned off, we observe good agreement with lattice data concerning the critical temperature, while the mass spectrum is of course nonphysical. If the anomaly is turned on, the critical temperature gets unreasonably high, but the mass spectrum is in good agreement with experimental data. This points out that explicit quark degrees of freedom and/or the Polyakov loop have to be introduced to get a lower value for $T_c$ with the spectrum remaining physical. These -- together with investigating the effect of the instanton-induced temperature dependence of the `t Hooft parameter, and extending the results to finite density -- represent future studies to be reported in the near future.
\vspace{0.7cm}
\section*{Acknowledgements}
This work was partially supported by the Grant-in-Aid for Scientific Research (C) No. JP26400273.

\makeatletter
\@addtoreset{equation}{section}
\makeatother 

\renewcommand{\theequation}{A\arabic{equation}} 

\begin{widetext}

\appendix
\section{Field derivatives of the effective potential}

The ground state of the system belongs to the minimum of the effective potential (\ref{Eq:AnsatzV}) at $k\rightarrow 0$. One obtains the corresponding $v_0$ and $v_8$ condensates by requiring the respective first derivatives to be zero (the others are identically zero in a background of $v_0,v_8$):
\begin{subequations}
\bea
\frac{\partial V_{k}}{\partial s_0}\bigg|_{v_0,v_8}=\Big(U_k'(I_1)+C_k'(I_1)+A_k'(I_1)\Big)\frac{\partial I_1}{\partial s_0}\bigg|_{v_0,v_8}+C_k(I_1)\frac{\partial I_2}{\partial s_0}\bigg|_{v_0,v_8}+A_k(I_1)\frac{\partial I_{\det}}{\partial s_0}\bigg|_{v_0,v_8}+h_0 \equiv 0, \\
\frac{\partial V_{k}}{\partial s_8}\bigg|_{v_0,v_8}=\Big(U_k'(I_1)+C_k'(I_1)+A_k'(I_1)\Big)\frac{\partial I_1}{\partial s_8}\bigg|_{v_0,v_8}+C_k(I_1)\frac{\partial I_2}{\partial s_8}\bigg|_{v_0,v_8}+A_k(I_1)\frac{\partial I_{\det}}{\partial s_8}\bigg|_{v_0,v_8}+h_8 \equiv 0.
\eea
\end{subequations}
The required field derivatives of invariants are as follows:
\bea
\frac{\partial I_1}{\partial s_0}\bigg|_{v_0,v_8}&=&v_0, \qquad \frac{\partial I_1}{\partial s_8}\bigg|_{v_0,v_8}=v_8, \qquad
\frac{\partial I_2}{\partial s_0}\bigg|_{v_0,v_8}=v_8^2\Big(\frac{2v_0}{3}-\frac{1}{3\sqrt{2}}v_8\Big), \qquad
\frac{\partial I_2}{\partial s_8}\bigg|_{v_0,v_8}=v_8\Big(\frac{2v_0^2}{3}-\frac{v_0v_8}{\sqrt{2}}+\frac{v_8^2}{6}\Big), \nonumber\\
\frac{\partial I_{\det}}{\partial s_0}\bigg|_{v_0,v_8}&=&\frac{2v_0^2-v_8^2}{2\sqrt{6}}, \qquad
\frac{\partial I_{\det}}{\partial s_8}\bigg|_{v_0,v_8}=-\frac{v_8(\sqrt{2}v_0+v_8)}{2\sqrt{3}}.
\eea
Similarly, the masses of the scalar and pseudoscalar particles needed to evaluate (\ref{Eq:effpotflow}) are the following:
\begin{subequations}
\bea
m_{s,ij}^2\equiv \frac{\partial^2 V_k}{\partial s_i \partial s_j}&=&\delta_{ij}\Big(U_k'(I_1)+ I_2C_k'(I_1)+A_k'(I_1) I_{\det}\Big)+\frac{\partial^2 I_2}{\partial s^i\partial s^j}C_k(I_1)+\frac{\partial^2 I_{\det}}{\partial s^i \partial s^j}A_k(I_1) \nonumber\\
&+&\frac{\partial I_1}{\partial s^i}\frac{\partial I_1}{\partial s^j}\Big(U_k''(I_1)+I_2C_k''(I_1)+A_k''(I_1) I_{\det}\Big)
+\Big(\frac{\partial I_1}{\partial s^i}\frac{\partial I_2}{\partial s^j}+\frac{\partial I_1}{\partial s^j}\frac{\partial I_2}{\partial s^i}\Big)C_k'(I_1)\nonumber\\
&+&\Big(\frac{\partial I_1}{\partial s^i}\frac{\partial I_{\det}}{\partial s^j}+\frac{\partial I_1}{\partial s^j}\frac{\partial I_{\det}}{\partial s^i}\Big) A_k'(I_1),
\\
m_{\pi,ij}^2\equiv \frac{\partial^2 V_k}{\partial \pi_i \partial \pi_j}&=&\delta_{ij}\Big(U_k'(I_1)+ I_2C_k'(I_1)+A_k'(I_1) I_{\det}\Big) +\frac{\partial^2 I_2}{\partial \pi^i\partial \pi^j}C_k(I_1)+\frac{\partial^2 I_{\det}}{\partial \pi^i \partial \pi^j}A_k(I_1),
\eea
\end{subequations}
where we have already used that all first field derivatives with respect to $\pi_i$ are zero in a background field defined by $v_0,v_8$. The remaining nonzero derivatives are
\begin{subequations}
\bea
\frac{\partial^2 I_2}{\partial s_i s_j}\bigg|_{v_0,v_8}&=&
\begin{cases}
\frac{2}{3}v_8^2,  \hspace{4.5cm} \ife \hspace{0.1cm} i=j=0\\
-\frac{v_8^2}{\sqrt{2}}+\frac{4}{3}v_0v_8,  \hspace{3.0cm} \ife \hspace{0.1cm} i=0,\hspace{0.1cm} j=8 \hspace{0.1cm} \orr \hspace{0.1cm} i=8,\hspace{0.1cm} j=0\\
\frac{2}{3}v_0^2+\frac{v_8^2}{2}-\sqrt{2}v_0v_8,  \hspace{2.1cm} \ife \hspace{0.1cm} i=j=8\\
\frac{2}{3}v_0^2+\frac{v_8^2}{6}+\sqrt{2}v_0v_8, \hspace{2.1cm} \ife \hspace{0.1cm} i=j=1,2,3\\
\frac{2}{3}v_0^2+\frac{v_8^2}{6}-\frac{1}{\sqrt{2}}v_0v_8, \hspace{2.15cm} \ife \hspace{0.1cm} i=j=4,5,6,7\\
0, \hspace{4.9cm}  \els\\
\end{cases}\\
\frac{\partial^2 I_2}{\partial \pi^i \pi^j}\bigg|_{v_0,v_8}&=&
\begin{cases}
0, \hspace{4.9cm} \ife \hspace{0.1cm} i=j=0\\
-\frac{v_8^2}{3\sqrt{2}}+\frac{2}{3}v_0v_8, \hspace{2.9cm} \ife \hspace{0.1cm} i=0,\hspace{0.1cm} j=8 \hspace{0.1cm} \orr \hspace{0.1cm} i=8,\hspace{0.1cm} j=0\\
\frac{v_8^2}{6}-\frac{\sqrt{2}}{3}v_0v_8, \hspace{3.18cm} \ife \hspace{0.1cm} i=j=8\\
-\frac{v_8^2}{6}+\frac{\sqrt{2}}{3}v_0v_8, \hspace{2.88cm} \ife \hspace{0.1cm} i=j=1,2,3\\
\frac{5}{6}v_8^2-\frac{1}{3\sqrt{2}}v_0v_8, \hspace{2.83cm} \ife \hspace{0.1cm} i=j=4,5,6,7\\
0, \hspace{4.9cm}  \els\\
\end{cases}
\eea
\end{subequations}
and furthermore,
\begin{subequations}
\bea
\frac{\partial^2 I_{\det}}{\partial s_i s_j}\bigg|_{v_0,v_8}&=&
\begin{cases}
\sqrt{\frac23}v_0,  \hspace{7.0cm} \ife \hspace{0.1cm} i=j=0\\
-\frac{v_8}{\sqrt6},  \hspace{7.2cm} \ife \hspace{0.1cm} i=0,\hspace{0.1cm} j=8 \hspace{0.1cm} \orr \hspace{0.1cm} i=8,\hspace{0.1cm} j=0\\
-\frac{v_0}{\sqrt6}-\frac{v_8}{\sqrt3},  \hspace{6.3cm} \ife \hspace{0.1cm} i=j=8\\
-\frac{v_0}{\sqrt6}+\frac{v_8}{\sqrt3}, \hspace{6.3cm} \ife \hspace{0.1cm} i=j=1,2,3\\
-\frac{v_0}{\sqrt6}-\frac{v_8}{2\sqrt3}, \hspace{6.15cm} \ife \hspace{0.1cm} i=j=4,5,6,7\\
0, \hspace{7.7cm}  \els\\
\end{cases}\\
\frac{\partial^2 I_{\det}}{\partial \pi_i \pi_j}\bigg|_{v_0,v_8}&=&
\begin{cases}
-\sqrt{\frac23}v_0,  \hspace{6.75cm} \ife \hspace{0.1cm} i=j=0\\
\frac{v_8}{\sqrt6},  \hspace{7.45cm} \ife \hspace{0.1cm} i=0,\hspace{0.1cm} j=8 \hspace{0.1cm} \orr \hspace{0.1cm} i=8,\hspace{0.1cm} j=0\\
\frac{v_0}{\sqrt6}+\frac{v_8}{\sqrt3},  \hspace{6.55cm} \ife \hspace{0.1cm} i=j=8\\
\frac{v_0}{\sqrt6}-\frac{v_8}{\sqrt3}, \hspace{6.55cm} \ife \hspace{0.1cm} i=j=1,2,3\\
\frac{v_0}{\sqrt6}+\frac{v_8}{2\sqrt3}, \hspace{6.4cm} \ife \hspace{0.1cm} i=j=4,5,6,7\\
0. \hspace{7.75cm}  \els
\end{cases}
\eea
\end{subequations}

\end{widetext}


\begin{thebibliography}{9}
\bibitem{thooft76} G. `t Hooft, Phys. Rev. D {\bf 14}, 3432 (1976).
\bibitem{gross81} D. J. Gross, R. D. Pisarski, and L. G. Yaffe, Rev. Mod. Phys. {\bf 53} (1981).
\bibitem{fukushima13} K. Fukushima and C. Sasaki, Prog. Part. Nucl. Phys. {\bf 72} 99 (2013).
\bibitem{vertesi11} R. Vertesi, T. Csorgo, and J. Sziklai, Phys. Rev. C {\bf 83}, 054903 (2011).
\bibitem{mitter14} M. Mitter and B.-J. Schaefer, Phys. Rev. D {\bf 89}, 054027 (2014).
\bibitem{cossu15} G. Cossu, H. Fukaya, S. Hashimoto, J. Noaki, and A. Tomiya, arXiv:1511.05691.
\bibitem{dick15} V. Dick, F. Karsch, E. Laermann, S. Mukherjee, and S. Sharma, Phys. Rev. D {\bf 91}, 094504 (2015).
\bibitem{sharma16} S. Sharma, V. Dick, F. Karsch, E. Laermann, and S. Mukherjee, in {\it Proceedings of Quark Matter 2015, Kobe, Japan, September 27-October 3, 2015} (to be published),  arXiv:1602.02197.
\bibitem{kunihiro89} T. Kunihiro, Phys. Lett. B {\bf 219}, 363 (1989); Phys. Lett. B {\bf 245}, 687(E) (1990). 
\bibitem{fukushima01} K. Fukushima, K. Ohnishi and K. Ohta, Phys. Rev. C {\bf 63}, 045203 (2001).
\bibitem{nagahiro06} H. Nagahiro, M. Takizawa and S. Hirenzaki, Phys. Rev. C {\bf 74}, 045203 (2006).
\bibitem{nanova15} M. Nanova, EPJ Web Conf.\  {\bf 97}, 00022 (2015).  
\bibitem{LEPS} LEPS, http://www.lns.tohoku.ac.jp/~hadron/bgoegg.html.
\bibitem{fejos15} G. Fejos, Phys. Rev. D {\bf 92}, 036011 (2015).
\bibitem{fukushima11} K. Fukushima and T. Hatsuda, Rep. Prog. Phys. {\bf 74}, 014001 (2011).
\bibitem{pisarski84} R. D. Pisarski and F. Wilczek, Phys. Rev. D {\bf 29}, 338 (1984).
\bibitem{grahl13} M. Grahl and D.-H. Rischke, Phys. Rev. D {\bf 88}, 056014 (2013).
\bibitem{pelissetto13} A. Pelissetto and E. Vicari, Phys. Rev. D {\bf 88}, 105018 (2013).
\bibitem{grahl14} M. Grahl, Phys. Rev. D {\bf 90}, 117904 (2014).
\bibitem{nakayama14} Y. Nakayama and T. Ohtsuki, Phys. Rev. D {\bf 91}, 021901 (2015).
\bibitem{parganlija13} D. Parganlija, P. Kovacs, Gy. Wolf, F. Giacosa, and D.-H. Rischke, Phys. Rev. D {\bf 87}, 014011 (2013).
\bibitem{eser15} J. Eser, M. Grahl, and D.-H. Rischke, Phys. Rev. D {\bf 92}, 096008 (2015).
\bibitem{fejos14} G. Fejos, Phys. Rev. D {\bf 90}, 096011 (2014).
\bibitem{wetterich93} C. Wetterich, Phys. Lett. B {\bf 301}, 90 (1993).
\bibitem{berges02} J. Berges, N. Tetradis, and C. Wetterich, Phys. Rep. {\bf 363}, 223 (2002).
\bibitem{pawlowski07} J. M. Pawlowski, Ann. Phys. {\bf 322}, 2831 (2007).
\bibitem{skokov10} V. Skokov, B. Stokic, B. Friman, and K. Redlich, Phys. Rev. C {\bf 82}, 015206 (2010).
\bibitem{herbst11} T. K. Herbst, J. M. Pawlowski, and B.-J. Schaefer, Phys. Lett. B {\bf 696}, 58 (2011).
\bibitem{herbst13} T. K. Herbst, J. M. Pawlowski, and B.-J. Schaefer, Phys. Rev. D {\bf 88}, 014007 (2013).
\bibitem{herbst14} T. K. Herbst, M. Mitter, J. M. Pawlowski, B.-J. Schaefer, and R. Stiele, Phys. Lett. B {\bf 731}, 248 (2014).
\bibitem{tripolt14} R.-A. Tripolt, N Strodthoff, L. von Smekal, and J. Wambach, Phys. Rev. D {\bf 89}, 034010 (2014).
\bibitem{heller15} M. Heller and M. Mitter, arXiv:1512.05241.
\bibitem{pawlowski98} J. M. Pawlowski, Phys. Rev. D {\bf 58}, 045011 (1998).
\bibitem{jiang12} Y. Jiang, and P. Zhuang, Phys. Rev. D {\bf 86}, 105016 (2012).
\bibitem{kamikado15} K. Kamikado and T. Kanazawa, J. High Energy Phys. 01 (2015) 129.
\bibitem{fejos13} G. Fejos, Phys. Rev. D {\bf 87}, 056006 (2013).
\bibitem{litim01} D. F. Litim, Phys. Rev. D {\bf 64}, 105007 (2001).
\bibitem{pdg} Particle Data Group, http://pdglive.lbl.gov/Viewer.action
\bibitem{peskin} M. E. Peskin, and D. V. Schroeder, {\t An Introduction to Quantum Field Theory} (Westview Press, Boulder, CO, 1995).
\bibitem{heinz12} A. Heinz, F. Giacosa, and D.-H. Rischke, Phys. Rev. D {\bf 85}, 056005 (2012).

\end{thebibliography}
\end{document}